\documentstyle[prl,aps,multicol]{revtex}
\tighten
\input psfig
\begin{document}
\draft
\title{Dephasing and the 
Orthogonality Catastrophe in Tunneling through a Quantum Dot: 
the ``Which Path?'' Interferometer}
\author{I.L. Aleiner$^{1}$, Ned S. Wingreen$^{1}$, and Yigal Meir$^{2}$}
\address{$^{1}$NEC Research Institute,
4 Independence Way, Princeton, NJ 08540\\ $^{2}$
 Physics Department, Ben Gurion University, Beer Sheva, 84105, 
Israel
}
\maketitle

\begin{abstract}

The ``Which Path?'' interferometer consists of an Aharonov-Bohm
ring with a quantum dot (QD) built in one of its arms, and an additional 
quantum point contact (QPC) located close to the QD.
The transmission coefficient of the QPC 
depends on the charge state of the   
QD. Hence the point contact causes
controllable  dephasing  of transport through the QD,
and acts as a measurement device
for which path an electron takes through the ring.
We calculate the suppression of the Aharonov-Bohm oscillations which
is caused both by 
dephasing and by the orthogonality catastrophe, {\it i.e.},
respectively, by 
real and virtual electron-hole pair creation at the 
QPC. 
\end{abstract}

\begin{multicols}{2}

The interference between different trajectories of a particle is one 
of the central postulates of quantum mechanics. The transition
between classical and quantum behavior depends on when and whether
this interference is realized. With the advent of mesoscopic 
conducting structures, it has become possible  
to study directly the coherence between different trajectories
of an electron in a metal or 
semiconductor Aharonov-Bohm ring.
Among the phenomena observed in these systems are 
Universal conductance fluctuations, 
weak localization, and inelastic dephasing by
electron-electron and electron-phonon scattering\cite{review}.
Recently, a set of elegant Aharonov-Bohm ring experiments 
was performed to detect
the phase shift of electrons passing through a quantum 
dot (QD) built in one arm of the ring\cite{yacoby,schuster}.
These experiments were the first to
demonstrate the coherent propagation of electrons through a quantum dot. 

The observation of phase coherence in transport through a QD
presents an opportunity to study the origins
of {\it decoherence} in mesoscopic structures.   
Recent work in atomic physics
has measured decoherence rates of the electromagnetic field
in a cavity\cite{haroche}. These experiments, however,
did not  control the rate of dephasing.
An Aharonov-Bohm ring with a QD in one of its arms,
offers the ability not only to measure dephasing rates, 
but to directly control
these rates by modifying the environment of the quantum system.
The proposed experimental set up for this ``Which Path?''
interferometer\cite{experiment} is shown in Fig.~\ref{fig}. 
An electron traversing the ring may follow the upper or the
lower arm. In the latter case, the electron must pass through
a QD located in the lower arm. In the proposed experiment, 
an additional wire containing a quantum point
contact (QPC) is placed close to the QD. 
The electrostatic field of an extra electron on the QD 
changes the transmission coefficient ${\cal T}$
of the nearby QPC, 
and hence changes the conductance of the wire. 
The change in the current in the wire 
``measures''  
which path the electron took around the ring,
causes the paths to decohere, and so  
suppresses the Aharonov-Bohm oscillations. 


{\narrowtext
\begin{figure}[h]
\vspace{0.5cm}
\hspace*{-0.20cm}
\psfig{figure=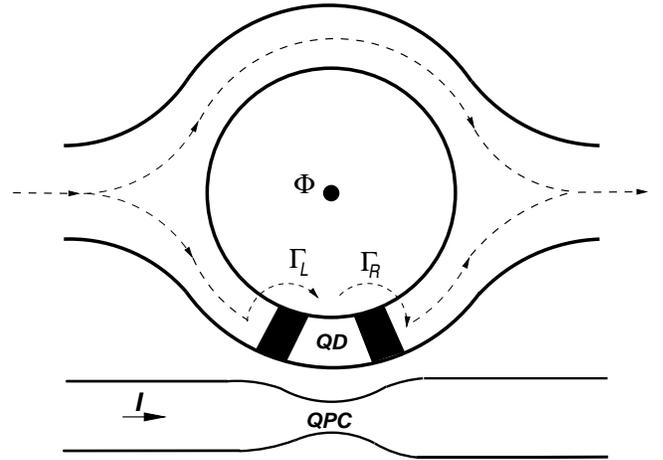,height=6.10cm}
\vspace{0.7cm}
\caption{Schematic view of the ``Which Path?'' interferometer
\protect\cite{experiment}.  The quantum dot (QD) is built in the lower
arm of an Aharonov-Bohm ring, as shown. The transmission coefficient
of the nearby quantum point contact (QPC) depends on the occupation
number of the dot because of electrostatic interactions.
(Four-terminal measurement is implied, so that closed orbits in
the ring are not important.)  }
\vspace{-0.22cm}
\label{fig}
\end{figure}
}

To estimate the rate of decoherence induced by the current in the wire    
consider the following  argument:
Adding an electron to the dot changes the conductance of the
QPC by $2 (e^2/h) \Delta {\cal T}$.
Detection of this electron requires a time $t_d$
such that the  change in the number of electrons crossing
the QPC exceeds the typical quantum shot noise,
\begin{equation}
t_d {V \over e } { {2 e^2} \over h} 
\Delta {\cal T}
\geq \sqrt{t_d {V \over e } 
 { {2 e^2} \over h} {\cal T}(1 - {\cal T}) },
\label{tdone}
\end{equation}
where $V$ is the bias voltage in the wire, and the right hand
side reflects the quantum shot noise across the QPC\cite{lesovik}.
The decoherence rate, therefore,  depends on both the bias across the QPC
and its transmission coefficient: 
\begin{equation}
{1 \over {t_d}} \approx  
{ {eV} \over h}
{ {(\Delta {\cal T})^2} \over 
{{\cal T} (1 - {\cal T})} }.
\label{tdtwo}
\end{equation}

In this paper, we calculate nonperturbatively 
the suppression of the Aharonov-Bohm
oscillations in a ring with a QD 
due to the close proximity of a wire containing a QPC.
 Our results support the simple argument given above,
and explicitly
show that $1/t_d$ is the rate of real electron-hole
pair creation in the wire. The simple estimate (\ref{tdtwo}) however
neglects the effect of virtual electron-hole pairs. 
The latter do not directly cause
decoherence but they decrease
the transmission amplitude through the QD. These
virtual processes result in {\it power-law suppression} of the 
Aharonov-Bohm oscillations.
This is an example of the 
orthogonality catastrophe \cite{anderson,mahan}, 
and is an inevitable consequence of ``measurement" by local
interaction with a many-body system. (We neglect the additional 
orthogonality catastrophe due to ring electrons\cite{matveev} 
because it cannot be externally controlled.)  

In the proposed experiment, the transmission coefficient across
the ring ${\cal T}_{\rm ring}$ can be
obtained from the appropriate combination
of measurements in a multiprobe geometry\cite{schuster}.
According to the Aharonov-Bohm effect,  {\it i.e.}, the phase 
difference of $2\pi \Phi/\Phi_0$
between electron trajectories which encompass a 
magnetic flux $\Phi$, one has 
\begin{equation}
{\cal T}_{\rm ring} = {\cal T}_{\rm ring}^{(0)} +
{\rm Re}\{ t^*  t_{\scriptscriptstyle{QD}} e^{2\pi i \Phi/\Phi_0} \}
+ \dots,
\label{sigmaphi}
\end{equation}
where the dots indicate higher harmonics of $\Phi$, and $\Phi_0 =
hc/e$ is the flux quantum.  The magnetic-flux independent term ${\cal
T}_{\rm ring}^{(0)}$ and the amplitude $t^*$ are sensitive to the
geometry of the system, ({\em e.g.} the structure of the leads,
lengths of the arms, etc.).  The amplitude
$t_{\scriptscriptstyle{QD}}$ for coherent transmission through the dot
reflects only the properties of the dot and its immediate environment;
this quantity will be discussed in the remainder of this paper.

We are interested in the Aharonov-Bohm oscillations
in the vicinity of Coulomb blockade peaks, {\it i.e.}, near the charge
degeneracy point  of the QD. 
This means, that only
two charging states of the dot, $N$ and $N+1$, are relevant
to transport\cite{qdot}. 
We neglect energy dependence of the phase from propagation
down the arms of the ring\cite{ring}, so that
$t_{\scriptscriptstyle{QD}}=\int\! d\epsilon
(-\partial f/\partial \epsilon)\, t_{\scriptscriptstyle{QD}}(\epsilon)$,
where $f(\epsilon)$
is the
Fermi distribution function (all energies are counted from the
Fermi level) and $t_{\scriptscriptstyle{QD}}(\epsilon)$ 
is the transmission amplitude for an electron
with energy $\epsilon$ through the QD.

In the Coulomb-blockade regime the broadening of levels is smaller 
than the level spacing in the dot\cite{qdot}. 
Thus, it is natural to consider only a single resonant
level in the dot.
The amplitude $t_{\scriptscriptstyle{QD}}(\epsilon)$ can then
be expressed in terms
of the exact retarded Green function of this level:
\begin{equation}
t_{\scriptscriptstyle{QD}}(\epsilon) = -i\sqrt{4\Gamma_L\Gamma_R}
\int\! dt\, e^{i\epsilon t} G^r (t),
\label{g1}
\end{equation}
where $\Gamma_{L,R}$ are the widths of the  level with
respect to tunneling to the left or to the right. The retarded Green
function is defined as
$G^r(t)=-i\theta (t)\langle\hat{c} (t)\hat{c}^\dagger(0)
+ \hat{c}^\dagger (0)\hat{c}(t) \rangle$, where $\hat{c}(t)$ is the 
Heisenberg operator which removes an
electron from the resonant level (we
put $\hbar =1$). 

The electrons in the  dot 
interact with the electrons in the wire. Only the local scattering 
potential of the QPC is significantly affected
by this electrostatic interaction.
We use the standard description of a
QPC\cite{qpc}  as a 1D noninteracting
electron system, and choose the basis of scattering eigenstates 
corresponding to the potential in the QPC when exactly $N$
electrons  occupy the QD:
\begin{equation}
\hat{H}_N=\int \frac{dk}{2\pi}k\left(\psi^\dagger_{\cal
L}(k)\psi_{\cal L}(k)+\psi^\dagger_{\cal R}(k)\psi_{\cal R}(k) \right).
\label{H0}
\end{equation}
$\psi_{{\cal L},{\cal R}}$ are the fermionic operators for the
scattering states moving from the left and right, respectively, 
with summation over spin indices implied.
We linearize the spectrum and put the Fermi
velocity in the wire  $v_F =1$. 
The electrostatic
field of an additional ($N+1{\rm st}$) electron on the QD 
changes the wire Hamiltonian to 
$\hat{H}_{N+1} = \hat{H}_N + \hat{V}$:
\begin{eqnarray}
&&\hat{V}(t)= \hat{V}_\phi(t) + \hat{V}_\chi (t); \label{U}\\ 
&&\hat{V}_\phi(t)=\lambda_\phi\int\frac{dk_1dk_2}{2\pi}\left(\psi^\dagger_{\cal
L}(k_1,t)\psi_{\cal L}
(k_2,t)+{\cal L}\leftrightarrow {\cal R} \right), \nonumber \\
&&\hat{V}_\chi(t)=\lambda_\chi\int\frac{dk_1dk_2}{2\pi}\left(\psi^\dagger_{\cal
L}(k_1,t)\psi_{\cal R}(k_2,t)e^{ieVt}
+ {\rm H.c.}\right) \nonumber
\end{eqnarray}
where the $\hat{\psi}(t) = e^{i\hat{H}_0t}\hat{\psi}e^{-i\hat{H}_0t}$
are electron operators in the interaction representation.  The
operator $\hat{V}_\phi(t)$ mixes scattering states propagating in the
same direction and corresponds to a change in the phase of the
transmission amplitude of the QPC.  The mixing between scattering
states which are incident from opposite directions is given by
$\hat{V}_\chi(t)$, and corresponds to a change in the transmission
coefficient $\cal{T}$ of the QPC. The explicit oscillatory time
dependence of $\hat{V}_\chi(t)$, describes a finite bias in the wire, {\it
i.e.}, $eV$ corresponds to the chemical potential difference between
${\cal L}$ and ${\cal R}$ scattering states.

The
Green function of the resonant level in the dot interacting with the
wire can be
approximated as
\begin{equation}
G^r(t)=-i \theta (t)e^{-i\epsilon_0 t -{\Gamma} t}
\left[P_{N+1}A_-(t) + P_{N} A_+(t)\right],
\label{onelevel}
\end{equation}
where $\epsilon_0$ is the single-electron energy of the level, and
$P_n$ is the probability of the corresponding charging state of the
dot, $P_N+P_{N+1}=1$. The total tunneling width of the level $\Gamma$
is given by $\Gamma = \Gamma_L +\Gamma_R$, and the coherence factors
$A_\pm(t)$ describe the influence of the wire,
\begin{mathletters}
\label{Apm}
\begin{eqnarray}
&&A_{+}(t)=\langle 
 e^{i\hat{H}_Nt}
 e^{-i\hat{H}_{N+1}t}
\rangle_{H_N},\\
\label{Ap}
&&A_{-}(t)=\langle 
 e^{i\hat{H}_Nt}
 e^{-i\hat{H}_{N+1}t}
\rangle_{H_{N+1}}.
\label{Am}
\end{eqnarray}
\end{mathletters}
The expectation values are taken with respect to an equilibrium
ensemble in the wire with the Hamiltonian, $H_N$ or $H_{N+1}$,
indicated as a subscript.  It is easy to see that Eq.~(\ref{onelevel})
is exact in two important limiting cases.  In the absence of the
interaction $A_\pm(t)=1$ and Eq.~(\ref{onelevel}) reduces to the retarded
Green function for a noninteracting resonant level, and Eq.~(\ref{g1})
becomes a simple Breit-Wigner formula.  Also, in the absence of
tunneling, $\Gamma=0$, Eqs.~(\ref{onelevel}) and (\ref{Apm}) are exact
expressions for an isolated level coupled to the wire. For the
intermediate regime $\Gamma > 0$, Eq.~(\ref{onelevel}) is not
exact. Physically, it neglects interaction induced correlations
between consecutive tunneling events of different electrons into
the dot. However, such events are rare in the case of weak tunneling,
and Eq.~(\ref{onelevel}) is expected to be a good approximation
even for $\Gamma \neq 0$.

Let us now turn to the calculation of the coherence factor $A_+(t)$.
For zero current in the wire, Eq.~(\ref{Apm}) corresponds to the well
known ``orthogonality catastrophe''\cite{anderson}, 
{\it i.e.}, the response of an equilibrium
noninteracting electron system to a sudden perturbation.
Exact results for this problem were first obtained in Ref.~\cite{nozieres}.
The long-time behavior ($eVt \gg 1$) of the nonequilibrium orthogonality 
catastrophe was recently considered by Ng\cite{ng}.
In order to find the dependence of $t_{\scriptscriptstyle{QD}}(\epsilon)$ 
on bias $eV$, we need
to know  $A_+(t)$ at all times. For the case of nonequilibrium in the wire
we were not able to obtain exact results for 
arbitrary constants $\lambda_{\phi,\chi}$. Instead,
we restrict ourselves to the case where the mixing between
scattering states is small, $\lambda_\chi \ll 1$, but 
$\lambda_\phi$ is arbitrary.

We begin by rewriting the coherence factor as
\begin{equation}
A_+(t)= \langle {\rm T}_t e^{-i\int_0^t\hat{V}(t_1)dt_1}\rangle_{H_N}=
A_\phi(t)A_\chi(t),
\label{aPP}
\end{equation}
 where $A_\phi(t)$ 
describes the orthogonality catastrophe in the absence of mixing 
between the scattering states:
\begin{equation}
A_\phi(t)=\langle{\rm T}_t e^{-i\int_0^tdt_1 \hat{V}_\phi(t_1)} \rangle_{H_N},
\label{App}
\end{equation}
 and can be evaluated exactly.
The results for the coherence factor (\ref{App}) 
are well-known\cite{nozieres}. One has
\begin{equation}
A_\phi (t) = 
\left(\frac{i\pi T}
{\xi_0\sinh \pi Tt}\right)^{4 \left(\frac{\delta_\phi}{\pi}\right)^2},\quad
\delta_\phi=\arctan \pi\lambda_\phi,
\label{App1}
\end{equation}
where $\xi_0$ is the high-energy cutoff, the smaller of the 
Fermi energy in the wire or the inverse rise time of the perturbation
of the QPC. The factor of four in the exponent in (\ref{App1})
corresponds to the number of affected channels 
(two scattering states multiplied by the spin degeneracy
in the wire). Equation (\ref{App1}) is identical to the expression 
describing  the ``shakeup''
effect in the X-ray absorption spectra in  metals\cite{nozieres},
which results in power-law
suppression $\epsilon^{4 \left({\delta_\phi}/{\pi}\right)^2}$ of the absorption
at low energies.

The factor $A_\chi(t)$ in (\ref{aPP}) describes the mixing of the scattering states 
in the wire and
we evaluate it in the linked-cluster approximation, keeping terms to
order $\lambda_\chi^2$:
\begin{equation}
A_\chi (t)=e^{-2\lambda_\chi^2
\int\!\int_0^tdt_1dt_2 \cos [eV(t_1-t_2)]g(t_1,t_2)g(t_2,t_1)}
\label{Ar}
\end{equation}
where the Green function $g(t_1,t_2)$ is defined as
\begin{eqnarray}
g(t_1,t_2)&=&-i A_\phi(t)^{-1}\label{g}\\
&\times & \int\!\frac{dk_1dk_2}{2\pi}\langle{\rm
T}_t\psi_1(k_1,t_1)\psi^\dagger_1(k_2,t_2)
 e^{-i\int_0^tdt \hat{V}_\phi(t)}\rangle.
\nonumber
\end{eqnarray}
The factor of two in the exponent in Eq.~(\ref{Ar}) 
comes from the summation over spin directions in the wire.
The Green function is given by\cite{nozieres}
\begin{eqnarray}
g(t_1,t_2)& =&
 \left(
\frac{\sinh \pi T (t-t_1)}
{\sinh \pi T(t-t_2)}\frac{\sinh
\pi T t_2}{\sinh \pi Tt_1}
\right)^{\frac{\delta_\phi}{\pi}}\label{gg1}\\
&\times & \left\{
 P \frac{\pi T\cos^2\delta_\phi }{\sinh \pi T(t_2-t_1)}
-\frac{\pi}{2} 
\delta(t_1-t_2) \sin 2\delta_\phi  \right\}
 \nonumber
\end{eqnarray}
where $P$ stands for the principal value, and $0 \leq t_{1,2} \leq t$.

Substituting $g(t_1,t_2)$ from 
Eq.~(\ref{gg1}) into Eq.~(\ref{Ar}), we obtain with the
help of Eq.~(\ref{aPP})
 \begin{equation}
A_+(t) =
\left(\frac{i\pi T}{\xi_0\sinh \pi Tt}\right)^{\alpha+\gamma }
e^{-\Gamma_d t+\gamma h\left(t,T,eV\right) },
\label{result1}
\end{equation}
where the exponents are related to
the scattering constants $\lambda_{\phi,\chi}$ from Eq.~(\ref{U}) by
\begin{equation}
\alpha = 4\left(\frac{\delta_\phi}{\pi}\right)^2,
\quad \gamma = 4\lambda_\chi^2\cos^4\delta_\phi,
\label{ind}
\end{equation}
and the dephasing rate is given by
\begin{equation} 
\Gamma_d = \pi\gamma|eV|.
\label{dephasing}
\end{equation}
The crossover function $h$ in Eq.~(\ref{result1}) is 
\[
h\left(t,T,eV\right)=
\int_0^{t}\! d\tau\, \tau (1-\cos eV\tau)\frac{\pi^2T^2}{\sinh^2\pi T \tau}.
\]

Let us now reexpress the exponents (\ref{ind}) in terms of the physical
characteristics of the QPC: the transmission probability
${\cal T}$ and the phase of the transmission amplitude $\theta$. In
order to do so, we notice that switching on the perturbation (\ref{U})
by adding an electron to the dot corresponds to changing the phase
shifts $\delta_{e,o}$ for the even $(e)$ and odd $(o)$ channels in the
wire:
\[
\delta_{e,o}^{(N+1)} = \delta_{e,o}^{(N)} + \Delta \delta_{e,o}, \quad
\Delta\delta_{e,o}= \arctan \pi (\lambda_\phi\pm\lambda_\chi).
\]
The transmission probability of the QPC is related to
these phase shifts by ${\cal T}=\cos^2(\delta_e - \delta_o)$, and
the phase of the transmission amplitude is given by $\theta =
\delta_e + \delta_o$. 
We obtain from Eq.~(\ref{ind})
\begin{eqnarray}
\alpha = \left(\frac{\Delta\theta}{\pi}\right)^2 + O(\lambda_\chi^2),
\ \ \ \ \ \gamma = \frac{\left (\Delta {\cal T}\right )^2}{8\pi^2 {\cal
T}\left(1-{\cal T} \right) }.
\label{index}
\end{eqnarray}
The dephasing rate $\Gamma_d$ given by
Eqs. (\ref{dephasing}) and (\ref{index}) agrees with the estimate for
$1/t_d$ obtained earlier in Eq.~(\ref{tdtwo}).

The physical meaning of the dephasing rate $\Gamma_d$ deserves some
additional discussion. Indeed, $\Gamma_d$ reflects the efficiency
with which the QPC measures the charge state of the quantum dot.
One can rigorously 
define this measurement using the
basis of scattering eigenstates of the wire before an electron is
added to the dot. If the added electron creates a 
single excitation in this basis, the passage of the electron through the 
QD is ``detected'' and interference with the
other, remote path through the Aharonov-Bohm ring is destroyed.
The dephasing rate $\Gamma_d$ is  the rate at which such
excitations are created.  Using the Golden Rule, for the
simplest case $\delta_\phi=0$, we obtain 
\[
{\Gamma_d} = 2\pi\lambda_\chi^2\cdot 2 \int_{-\infty}^0\!\!\!\!dk_i
\int_0^\infty \!\!\!dk_f \delta(k_i - k_f-|eV|) = 4\pi\lambda_\chi^2|eV|,
\]
which agrees with (\ref{dephasing}),
and which can easily be generalized to $ \delta_\phi \neq 0$.

Note the symmetry in the expressions for $\gamma$ and  $\Gamma_d$ 
 between the
transmission probability ${\cal T}$ and the reflection probability $1-
{\cal T}$ in the wire.  An extra electron transmitted through
a normally reflecting QPC provides the same measurement of
the charge state of the QD as an extra electron reflected by
a normally transmitting point contact. For the case of a   
parabolic potential barrier 
in the QPC, $\Delta {\cal T} \sim {\cal T}(1- {\cal T} ) 
\Delta V_{\scriptscriptstyle QPC}$, where 
 $\Delta V_{\scriptscriptstyle QPC}$ is the change in the height of
the potential caused by adding an electron to the  dot. One then finds
$\Gamma_d \propto {\cal T}(1- {\cal T} )$, with the maximum dephasing rate at
${\cal T} = 1/2$.


The calculation of the coherence factor $A_-(t)$ from Eq.~(\ref{Am})
is performed analogously, starting from the diagonalization
of the Hamiltonian $\hat{H}_{N+1}(t)$ in the basis of scattering
states.
The result is $A_-(t)=A_+(t)^*$. Because $A_-(t) \neq A_+(t)$, the
probability $P_N$ for the occupation of the dot does not cancel from
the result.
For the general position of the level $\epsilon_0$, the probability
$P_N$ can be found from the thermodynamic formula
$P_N = -\int ({d\epsilon}/{\pi})f(\epsilon){\rm Im}{G}^r(\epsilon)$. However,
at the peak of the 
Coulomb blockade, $\epsilon_0 = 0$, it is obvious that
$P_N = P_{N+1}=1/2$. The total transmission amplitude 
through the quantum dot can then be obtained (\ref{g1}) as a Fourier 
transform of $G^r(t)$. We find that the result can
be well approximated by the simple formula
\begin{equation}
t_{\scriptscriptstyle{QD}}\! 
\simeq \frac{2\sqrt{\Gamma_L\Gamma_R}}{T + \Gamma_{\rm tot} }
\left(\!\frac{ T + \Gamma_{\rm tot} }{\xi_0}\!\right)^\alpha\!
\left(\!\frac{ T + \Gamma_{\rm tot} + |eV|}{\xi_0}\!\right)^\gamma\!\!, 
\label{interpolation}
\end{equation}
where the total width is given by $\Gamma_{\rm tot}=\Gamma_L+\Gamma_R
+ \Gamma_d$ \cite{algebra}.
Equation (\ref{interpolation}) is the central result of
our study. It describes the anomalous scaling of the amplitude of the
Aharonov-Bohm oscillations with the temperature or with the current
flowing through the quantum wire. It is important to realize that
even at zero-temperature equilibrium the quantum wire suppresses the 
Aharonov-Bohm oscillations through the orthogonality catastrophe,
{\it i.e.} the suppression of tunneling through the quantum
dot by the creation of virtual electron-hole pairs in the wire.
In this limit, the conductance through the dot is 
$\sim |t_{\scriptscriptstyle{QD}} |^2$, so we expect 
the same power-law behavior as in 
(\ref{interpolation}) even  without the Aharonov-Bohm geometry.

In conclusion, we have analyzed theoretically 
electron transport through the ``Which Path?'' 
interferometer\cite{experiment}:
an Aharonov-Bohm ring with a quantum dot in one arm, 
and an additional wire containing a quantum
point contact located close to the dot.
The presence of the wire suppresses 
the Aharonov-Bohm oscillations in the ring in two ways.  
First, real electron-hole-pair creation
in the wire ``measures'' which path the electron took around the ring, 
and so causes the paths to decohere.
Second, virtual electron-hole-pair creation in the wire
decreases the transmission amplitude through the QD,
leading to power-law dependence of
the Aharonov-Bohm oscillations 
on the temperature or the current through the wire. 

We are thankful to L. Glazman, A. Larkin, A. Stern, and A.
Yacoby for valuable discussions. We also thank E. Buks and 
M. Heiblum, who, having learned of the present work, informed us
of their independent experiment and 
derivation of Eq.~(\ref{tdtwo})\cite{experiment}.
One of us (Y.M.) was partially
supported by the Israeli Science Foundation.

\end{multicols} 

\begin{references} 
\bibitem{review} See, {\it e.g.}, A.G. Aronov and Y.V. Sharvin, Rev.
Mod. Phys. {\bf 59}, 755 (1987) and references therein.
\bibitem{yacoby} A. Yacoby {\it et al.},
Phys. Rev. Lett., {\bf 74}, 4047 (1995); A. Yacoby, R. Schuster, and 
M. Heiblum, Phys. Rev. B {\bf 53}, 9583 (1996).
\bibitem{schuster} R.~Schuster {\it et al.},
Nature {\bf 385}, 417  (1997).
\bibitem{haroche} M. Brune 
{\it et al.}, Phys. Rev. Lett., {\bf 77}, 4887 (1996).
\bibitem{experiment} E. Buks and M. Heiblum, (unpublished).
We employ the name ` ``Which Path?" interferometer'
coined for their experimental structure 
even though our work was begun independently.
\bibitem{lesovik} G.B. Lesovik, JETP Lett. {\bf 49}, 591 (1989). 
\bibitem{anderson}P.W. Anderson, Phys. Rev. {\bf 164}, 352 (1967).
\bibitem{mahan} G. D. 
Mahan, {\it Many-Particle Physics} (Plenum, New York, 1990)
 2nd ed.
\bibitem{matveev} 
K.A. Matveev and A.I. Larkin, 
Phys. Rev. B {\bf 46}, 15337 (1992).
\bibitem{qdot}For a review, see U. Meirav and E. B. Foxman, 
Semicond. Sci. Technol. {\bf 10}, 255 (1995).
\bibitem{ring}This implies that the temperature
$T \ll\hbar v_F^{(\rm r)}/R$, where 
$v_F^{(\rm r)}$ is the Fermi velocity and $R$ is the radius
of the ring.
\bibitem{qpc} C. W. J. Beenakker and H. van Houten, 
Solid State Phys. {\bf 44}, 1 (1991).
\bibitem{nozieres} P. Nozieres and C.
T. De Dominicis, Phys. Rev. {\bf 178}, 1097 (1969).
\bibitem{ng} T.~K.~Ng, Phys. Rev. B {\bf 51}, 2009 (1995).
\bibitem{algebra} The dependence
of the amplitude of the Aharonov-Bohm oscillations on the dephasing
rate is only algebraic and not exponential. 
This is because the time an electron spends on the
QD  is  distributed
uniformly in the interval $\left[0,\Gamma^{-1}\right]$. 
The additional dephasing rate $\Gamma_d$ suppresses
contibutions from times larger than $1/\Gamma_d$, which leads to the
algebraic decay of the Aharonov-Bohm amplitude.

\end{references}
\end{document}